\documentclass[11pt]{article}
\usepackage[a4paper,top=2cm,bottom=2cm,left=2.5cm,right=2.5cm]{geometry}

\usepackage{mathptmx}
\usepackage[english]{babel}
\usepackage[utf8x]{inputenc}
\usepackage[T1]{fontenc}
\usepackage{ae,aecompl}
\usepackage{amsfonts}
\usepackage{amssymb}
\usepackage[numbers]{natbib}

\usepackage[fleqn]{amsmath}
\usepackage{graphicx}
\usepackage{subcaption}
\usepackage{url}
\usepackage[colorlinks=true, allcolors=blue]{hyperref}
\usepackage{authblk}

\begin{document}
\title{\textbf{Mode-Selective Laser Propagation and Absorption in Strongly Magnetized Inhomogeneous Plasma}}

\author{Kun Li$^{1,*}$, Wuhan Wu$^{1}$, Yuxi Li$^{1}$, Mingyang Yu$^{2}$}
\date{}
\maketitle

\begin{center}
\small
$^{1}$Department of Physics, College of Science, Shantou University, Shantou, Guangdong 515063, China\\
$^{2}$College of Engineering Physics, Shenzhen Technology University, Shenzhen 518118, China\\
$^{*}$Corresponding author. Email: kunli@stu.edu.cn
\end{center}

\vspace{12pt}
\textbf{Abstract}

\noindent We systematically investigate the field-aligned propagation and collisional absorption of normally incident laser light in a strongly magnetized inhomogeneous plasma. Analytical expressions for electric fields in both vacuum and plasma are derived. Using analytical modelling and particle-in-cell simulations, we establish the cutoff conditions, absorption efficiencies, and scaling laws for the right-hand (R) and left-hand (L) circularly polarized waves. The dependence of collisional absorption coefficient on magnetic field strength, plasma scale length and laser intensity are quantified. In particular, L waves reflect at  cutoff density,  with absorption strongly enhanced as the magnetic field increases. For the R-waves, the absorption decreases with increasing magnetic field when the normalized electron cyclotron frequency is less than unity. However, when it exceeds unity, the R-waves propagate as whistler modes without a cutoff, allowing penetration into overdense plasma. This enables deep energy deposition inside overdense plasma. These results provide a framework for understanding laser-plasma energy coupling through collisional absorption in strongly magnetized inhomogeneous plasma. 

\vspace{6pt}


\section{Introduction}

Over the last decade, advances in high-power lasers have enabled the generation of magnetic field reaching $10^2 \sim 10^3$ tesla in laboratories through methods such as laser-driven capacitor coils, snail targets and seed magnetic field compression \cite{fujioka2013kilotesla, ehret2022kilotesla, gotchev2009laser}. Particle-in-cell (PIC) simulations have shown magnetic field even up to $10^3 \sim 10^6$ tesla level \cite{perkins2013two,murakami2020generation,shi2023efficient,matsumura2023effect}. The extremely strong magnetic fields have opened new frontiers in the research of inertial confinement fusion (ICF) due to its influence on alpha-particle confinement, heat transport, laser-plasma coupling and et al. \cite{matsumura2023effect,chang2011fusion,moody2022magnetized}. 

The propagation of electromagnetic (em) waves in plasma along the magnetic field has been treated in many textbooks, with properties in cold plasma well summarized in the Clemmow–Mullaly–Allis diagram \cite{ginzburg1970propagation,krall1973principles,chen1985introduction}. For example, L wave reflects at L-cutoff ($n=1+B$), R wave reflects at R-cutoff ($B<1, ~n=1-B$) or propagates without cutoff ($B>1$, whistler wave). Here, $B=B_0/B_c$ and $n=n_e/n_c$, where $B_c=m_e c\omega/e \approx 1.07\times10^4 \lambda^{-1}$ tesla is the critical threshold of magnetic field, $n_c= m_e \omega^2/4\pi c \approx 1.12\times 10^{21}\lambda^{-2}cm^{-3}$ is the critical plasma density, with $m_e$, $c$, $e$, $\omega$, $\lambda$ being electron mass, light speed, electron charge, em frequency,  and em wavelength in unit of micrometers, respectively. Recent progresses in the generation of strong magnetic field have enabled the laser-plasma interactions to develop into the regime of highly-magnetized plasma ($0.01<B<100$) and have disclosed many interesting phenomena. For example, the interactions of relativistic ultrashort-pulse (USP) laser with highly-magnetized inhomogeneous plasma are investigated using Particle-in-Cell (PIC) simulation \cite{arefiev2016enhanced, yang2015propagation,wu2017controllable, gong2017enhancing, sano2024relativistic}, showing properties of enhanced plasma heating and particle acceleration, with whistle wave being more favourable. Based on analytical methods, the interactions of low- and moderate- intensity long-pulse laser with semi-infinite highly-magnetized homogeneous collisional plasma have been investigated \cite{li2020,li2022parametric}, showing properties such as plasma heating front and parametric dependence of energy deposition. 

However, a comprehensive investigation of laser interactions with highly-magnetized inhomogeneous collisional plasma remains unexplored. As a comparison with un-magnetized plasma,  the normal incidence and collisional absorption of electromagnetic wave or laser onto an inhomogeneous collisional plasma have been thoroughly studied and summarized in several textbooks \cite{kruer2003physics, atzeni2004physics, tikhonchuk2023particle}. For example, the electric fields of em wave in plasma with constant density gradient are already derived, although missing the reflectivity at vacuum-plasma interface \cite{kruer2003physics, tikhonchuk2023particle}. Dependences of collisional absorption coefficient on laser intensity and wavelength are also demonstrated for plasma with exponential density profile \cite{atzeni2004physics}. These investigations build a sound basic for the research of laser-plasma-interaction and applications such as ICF. This paper will follow similar routines to investigate the propagation and collisional absorption of laser in a highly-magnetized inhomogeneous collisional plasma. Section \ref{sec: SP} will simulate the propagation of a low-intensity USP laser into a highly-magnetized plasma; section \ref{sec: LP} will analytically study the equations of electric fields and the dependence of collisional absorption; section \ref{sec:discussion} will discuss the results and potential applications.  

\section{Propagation and collisional absorption}
\label{sec: propagation}
Propagation of a circularly polarized (CP) laser along the extremely strong magnetic field normally into an inhomogeneous collisional plasma will be investigated, where the right-hand circularly polarized (RCP) and left-hand circularly polarized (LCP) laser will arouse R- and L- waves, respectively. Several restrictions and approximations are made: (1)  the plasma is stationary and ions are immobile duration the propagation process; (2)  normal incidence excludes resonant absorption, making collisional absorption the dominant mechanism; (3) laser weakly perturbs electron motion, thus the electron oscillation energy driven by the laser is much smaller than the plasma thermal energy, $E_{osc} \ll E_{th}$, limiting the range of laser intensity. For example, the laser intensity is written as $I\lambda^{2}\approx 1.37*10^{18}a^2W/cm^{2}$ with $\lambda$ being laser wavelength in unit of micrometer,  $a=A/(m_ec^2/e) $ is the dimensionless laser potential vector and the oscillation energy of electrons driven by the non-relativistic laser is approximately to be $\epsilon_{osc} \approx 0.5 m_e a^2 c^2$ with $a \ll 1$. In this way, the oscillation energy of electrons is  
\begin{align}{\label{eq:oscillationEnergy}}
& E_{osc}(keV)\approx 0.5 m_e c^2 \times \frac{I \lambda^2 (W\cdot \mu\text{m}^2/cm^{2})}{1.37\times 10^{18}}\approx  0.2  \frac{I \lambda^{2}}{10^{15}W/cm^2}
\end{align}
Typical plasma thermal energy is in range of  $0.1\sim 5$ keV for laser-plasma interactions, thus requirement of $E_{osc} \ll E_{th}$ is equivalent to $I \ll 10^{15}~ W/cm^2$ with $\lambda=1~\mu\text{m}$ or $I \ll 10^{13}~ W/cm^2$ with $\lambda=10~\mu\text{m}$.  

The permittivity of collisional magnetized plasma for R- and L- wave is \cite{li2020}
\begin{align}
& \epsilon_{\pm}=1-\frac{n}{1\pm B+i\nu}
\label{eq:permittivity01}
\end{align}
Here, sign $-$ denotes R wave, sign $+$ denotes L wave. It is found \cite{li2020} that  $\nu=\nu_{ei}/\omega_L=n\nu_c$ is the dimensionless electron-ion collisional frequency, where $\nu_c \approx 1.72 ln \Lambda Z \lambda^{-1}T_{eV}^{-3/2}$ is the the dimensionless collisional frequency at critical density $n=1$,  with $Z$ being the free electron number per atom, $ln\Lambda$ being the Coulomb logarithm, $T_{eV}$ being the plasma temperature in electronvolts, $\lambda_L$ being the laser wavelength in micrometers. If $\nu \ll 1\pm B$ which is usually true for R and L wave not close to R-cutoff ($n=1-B$, $B<1$) and L-cutoff ($n=1+B$),  the real and imaginary parts of the complex refractive index  $\tilde{k}=\epsilon_{\pm}^{1/2}=k_r + ik_i$ are approximately to be
\begin{align}
& k_r\approx  \sqrt{1-\frac{n}{1\pm B}} \label{eq:kr}\\ 
& K = 2k_i \approx \frac{n^2 /(1\pm B)^{2}}{\sqrt{1-n/(1\pm B)}}\nu_c \label{eq:K}
\end{align}
here, $K$ is the spatial decay rate of laser energy. From the equation of permittivity, it is also found that L wave is strongly damped at L-cutoff and R wave is strongly damped at R-cutoff  in case $B<1$. 

\subsection{Short-Pulse Laser} 
\label{sec: SP}
\begin{figure*}[hbt!]
  \centering
     \includegraphics[width=\linewidth]{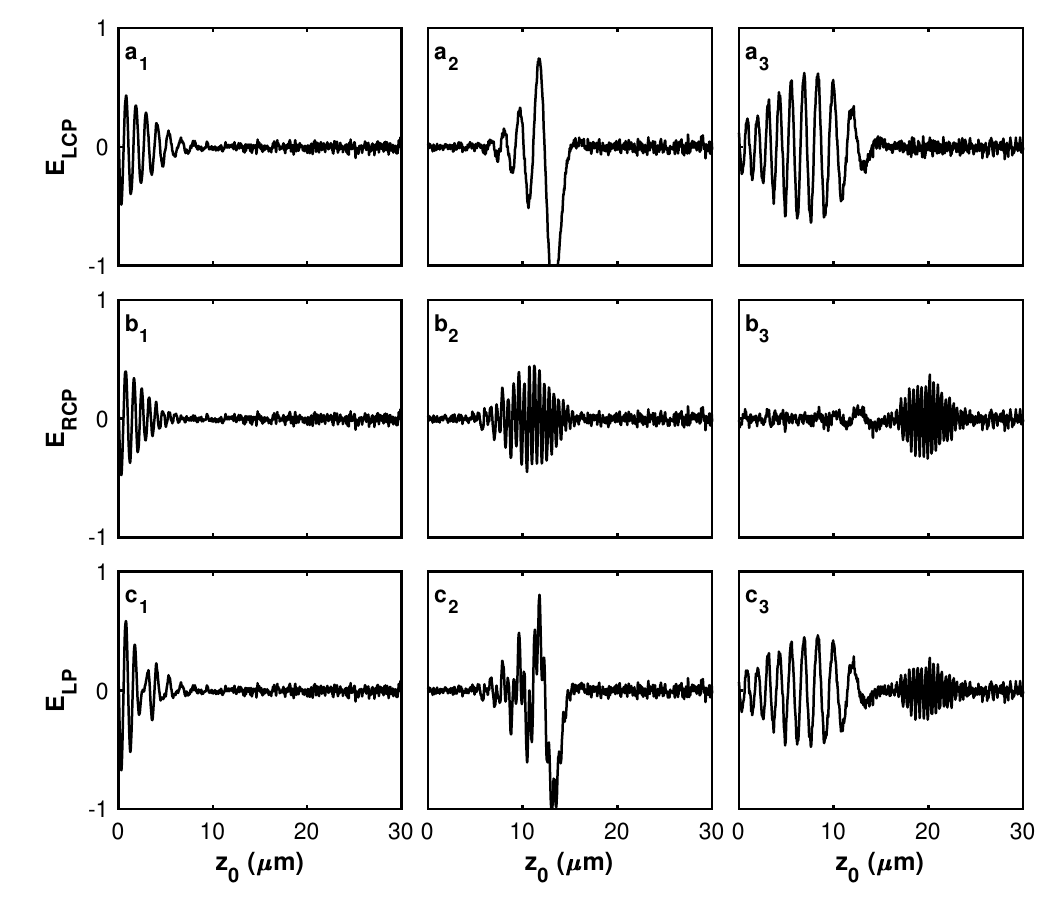}
   \caption{\label{fig:PIC} PIC simulation of short-pulse laser ($\lambda=1~\mu\text{m}$, pulse duration $30$ fs, $I_L=10^{14}~W/cm^{2}$) normally incident on a highly magnetized collisional plasma ($B=2$, $n_e=n_c \cdot z_0/5\mu\text{m}$, $T=100$ eV). ($a_1 \sim a_3$) LCP laser: $t=10$, 30 and 50 $T_0$ (left to right, same as below); ($b_1 \sim  b_3$) RCP laser; ($c_1 \sim c_3$) LP laser.}
\end{figure*}

The propagation process of an ultrashort shot-pulsed laser is investigated with 2D PIC simulations using EPOCH code, where the Gaussian RCP, LCP and linearly-polarized (LP)  lasers  ($1\mu{m}/30fs/10^{14}~ W/cm^2$) propagate normally into a plasma with constant density gradient ($n_e(z_0>0)=n_c \cdot z_0/L_0 $, $L_0=5~\mu\text{m}$, $T_{eV}=100$ eV)  along the magnetic field ($B_0=2\times 10^4$ tesla, $B=2$).  The simulation box has $70~ \mu\text{m} \times 30~ \mu\text{m}$ $2800\times 1200$ cells, with 40 cells per wavelength, 32 numerical macroparticles per species per cell, and simple boundary conditions for the fields. The laser electric fields at 10, 30 and 50 $T_0$ (laser period) are shown in Fig. \ref{fig:PIC} with following properties:
\begin{itemize}
\item LCP laser is red-shifted and reflects at $z_0=15~\mu\text{m}$ (L-cutoff) with enhanced amplitude;
\item RCP laser ($B>1$, whistler wave) is blue-shifted, pulse-compressed, without encountering cutoff and with much smaller amplitude compared to LCP laser at position of L-cutoff;
\item LP laser propagates with modulated profiles and is split into forward RCP and backward LCP pulses at L-cutoff.
\end{itemize}
These results are consistent with Eq.\eqref{eq:kr}. 

\subsection{Long-Pulse Laser} 
\label{sec: LP}

The propagation of a monochromatic laser  into a highly-magnetized collisional plasma with constant density gradient is investigated ($\lambda=1~\mu{m}$, $L_0=5~\mu{m}$  for most cases) with standing wave \cite{kruer2003physics,tikhonchuk2023particle} and travelling wave models \cite{atzeni2004physics} using dimensionless units,  i.e, $z=k z_0$, $t=\omega t_0$. 

\subsubsection{Standing wave model}
Standing wave model is employed to deduce the electric field profile and properties of collisional absorption but without accounting for heating effect, or in case of weak collision. The plasma density profile is $n(z<0)=0$ and $n(z \geq 0)=z/L$, thus permittivity of plasma is $\epsilon(z>0)=1-z/[L(1\pm B + i n \nu_c)]$. Assuming electric field of incident laser to be $E_{i,v} e^{-it} =E_0 exp(i(z-t))$,  the amplitude of whistler wave is 
\begin{align}{\label{eq:whistler}}
& E_{wh}(z \geq 0) = E_0 exp(i\tilde{k} z)
\end{align}
In vacuum, the electric field of reflected wave is $E_{r,v}(z<0)e^{-it}=r_0 E_0 exp(-i(z+t))$, thus the superposition of incident and reflected wave is $E_{s,v}(z<0)e^{-it}=E_{i,v}e^{-it}+E_{r,v}e^{-it}$; in plasma, in case of R-wave with $B<1$ or L-wave, the incident and reflected waves form a standing wave $E_{s,p}e^{-it}=E_{i,p}e^{-it}+E_{r,p}e^{-it}$. Both $E_{s,v}$ and $E_{s,p}$  satisfy the Helmholtz equation, and the equation for $E_{s,p}$ is,
\begin{align}
&\partial_{zz} E_{s,p} + \epsilon_{\pm} E_{s,p} =0
\label{eq:Helmholtz}
\end{align}
Since collisional absorption mainly takes place in region close to cutoff $n=1\pm B$,  the permittivity can be approximated to be $\epsilon \approx 1 - z/L_B$, $L_B \approx L(1\pm B)(1 + i\nu_c)$, as in the references \cite{kruer2003physics,tikhonchuk2023particle}. In this way, the Helmholtz equation has analytical solutions of Airy functions, $ Ai(\eta)$ and $Bi(\eta)$ with $\eta= L_B^{2/3} (z/L_B-1)$. Because $Bi(\eta)$ diverges to infinite with increased $z$, the physical solution is $\alpha Ai(\eta)$ with $\alpha$ being constant. In case of weakly inhomogeneous plasma $|L_B|>>1$, it is found that $ Ai(\eta)$ is approximately cosine function, and then the standing wave $E_{s,p}$, the incident wave $E_{i,p}$ and the reflected wave $E_{r,p}$ are expressed as
\begin{equation}
\begin{aligned}
&E_{s,p}(z\geq 0)  \approx \frac{\alpha}{\sqrt{\pi}(-\eta)^{1/4}} \cos\left[\frac{2}{3}(-\eta)^{3/2}-\frac{\pi}{4} \right]  \\
&E_{i,p} (z\geq 0) \approx \frac{\alpha}{2\sqrt{\pi}(-\eta)^{1/4}} exp \left[- i \left(\frac{2}{3}(-\eta)^{3/2}-\frac{\pi}{4} \right)\right]  \\
&E_{r,p} (z\geq 0) \approx \frac{\alpha}{2\sqrt{\pi}(-\eta)^{1/4}} exp \left[ i \left(\frac{2}{3}(-\eta)^{3/2}-\frac{\pi}{4} \right)\right]
\end{aligned}
\label{eq:plasma0}
\end{equation}
Due to the continuity of plasma density at vacuum-plasma interface $z=0$, the electric fields of incident and reflected laser are continuous respectively,  $E_{i,v}(z=0)=E_{i,p}(z=0)$, $E_{r,v}(z=0)=E_{r,p}(z=0)$ and  $\eta(0)=-L_B^{2/3}$. Then, the reflectivity of laser electric field at vacuum-plasma interface is
\begin{align}
\label{eq:reflectivity}
& r_0=\frac{E_{r,v}(z=0)}{E_{i,v}(z=0)} =exp\left[ i \left( \frac{4}{3}L_B -\frac{\pi}{2} \right) \right]
\end{align}
The coefficient of collisional absorption is
\begin{align}
\label{eq:absorption01}
& f_{abs}= 1-r_0r_0^*=1-exp\left[ -\frac{8}{3}L\left( 1\pm B\right)\nu_c \right]
\end{align}
And,
\begin{align}
&\alpha = E_0 \cdot 2\sqrt{\pi}L_B^{1/6}exp \left[ i \left(\frac{2}{3}L_B-\frac{\pi}{4} \right)\right]
\end{align}
Thus, the amplitudes of electric fields for incident, reflected and standing waves in plasma are
\begin{equation}
\begin{aligned}
& E_{i,p}(z>0)\approx \frac{E_0}{\left(1-z/L_B \right)^{1/4}} exp \left[ i \frac{2}{3}L_B \left( 1- \left(1-z/L_B\right) ^{3/2} \right)  \right] \\
& E_{r,p}(z>0)\approx \frac{E_0}{\left(1-z/L_B \right)^{1/4}} exp \left[ i \frac{2}{3}L_B \left( 1+ \left(1-z/L_B\right) ^{3/2} \right) -i \frac{\pi}{2} \right] \\
& E_{s,p}(z>0)=E_0 \cdot 2\sqrt{\pi}L_B^{1/6}exp \left[ i \left(\frac{2}{3}L_B-\frac{\pi}{4} \right)\right] Ai\left( \eta(z) \right)
\end{aligned}
\label{eq:plasma}
\end{equation}
In case of un-magnetized plasma, the electric fields and collisional absorption coefficient are obtained from above and are the same as results in the textbooks \cite{kruer2003physics,tikhonchuk2023particle}. 

\begin{figure*}[hbt!]
  \centering
    \begin{subfigure}[b]{0.49\linewidth}
    \includegraphics[width=\linewidth]{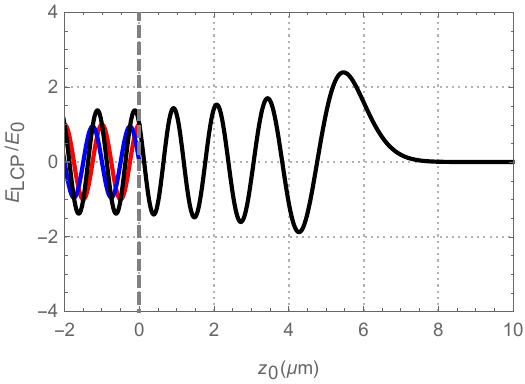}
    \caption{\label{fig:LCP1} }
    \end{subfigure}  
     \begin{subfigure}[b]{0.49\linewidth}
    \includegraphics[width=\linewidth]{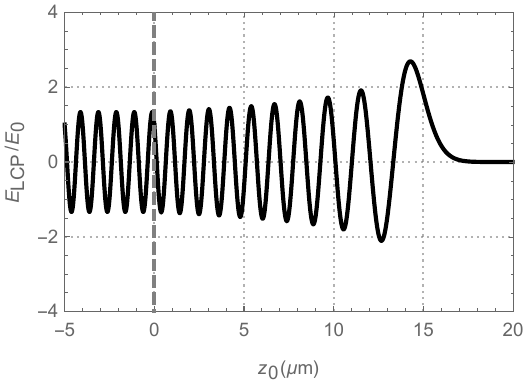}
    \caption{\label{fig:LCP2} }
   \end{subfigure}  
       \begin{subfigure}[b]{0.49\linewidth}
    \includegraphics[width=\linewidth]{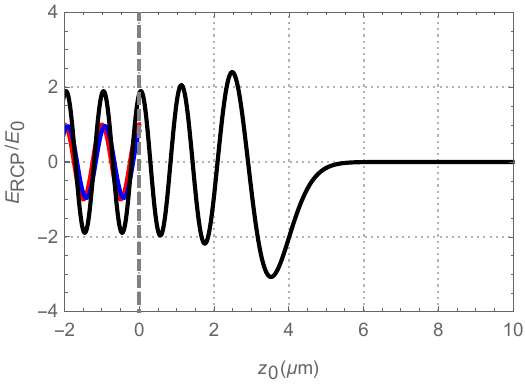}
    \caption{\label{fig:RCP1} }
   \end{subfigure}  
       \begin{subfigure}[b]{0.49\linewidth}
    \includegraphics[width=\linewidth]{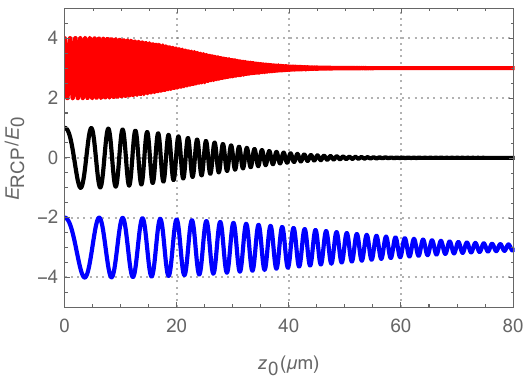}
    \caption{\label{fig:RCP2} }
   \end{subfigure}  
  \captionsetup{justification=raggedright,singlelinecheck=false}
  \caption{\label{fig:electricField} Relative electric field $E(z)/E_0$ from standing wave model ($\lambda=1~\mu\text{m}$, $n_e/n_c=z_0/L_0$  with $L_0=5~\mu\text{m}$, $\nu_c=0.001$).  (a) LCP with B=0.2: $E_{s,v}(z<0)=E_{i,v}+E_{r,v}$ (black) is the superposition of incident $E_{i,v}$ (red) and reflected laser $E_{r,v}$ (blue),  $E_{s,p}(z \geq 0)$ (black) is the standing wave in plasma;  (b) LCP, B=2:  $E_{s,v}$ and $E_{s,p}$; (c) RCP, $B=0.2$, similar to (a); (d) RCP (whistler mode): $B_0=2*10^4$ T and $\lambda=1~\mu{m}$ (red,  $B=2$), $B_0=6*10^3$ T and $\lambda=10~\mu{m}$ (black,  $B=6$), and $B_0=2*10^4$ T and $\lambda=10~\mu{m}$ (blue,  $B=20$), $L_0=20~\mu{m}$,  $\nu_c=0.01$.}
\end{figure*}

Based on above equations, the laser electric fields in both vacuum and plasma are obtained with $B=0.2, ~1.5, ~2$, as shown in Fig. \ref{fig:electricField}. Fig. \ref{fig:RCP2} also compares the propagation of $Nd:YAG$ laser and $CO_2$ laser in whistler mode with different magnetic field strength. It is found that even with weaker magnetic field, the long-wavelength laser can propagate as deep as shorter-wavelength laser with stronger magnetic field.

Dependences of collisional absorption efficient on plasma scale length and strength of magnetic field are also illustrated in Fig. \ref{fig:01} and \ref{fig:02}, respectively, showing that $f_{abs}$ increases with plasma scale length and magnetic field for L wave, while decreases with magnetic field for R wave ($B<1$). 

\subsubsection{Travelling wave model}
\begin{figure*}[hbt!]
  \centering
    \begin{subfigure}[b]{0.49\linewidth}
    \includegraphics[width=\linewidth]{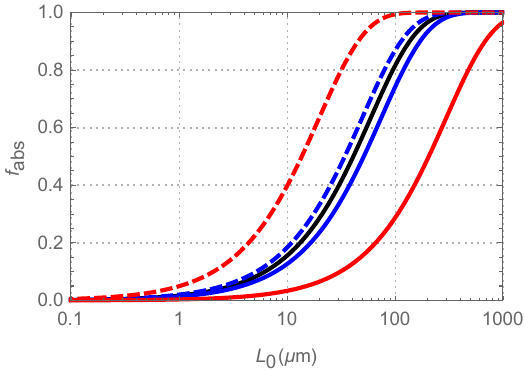}
    \caption{\label{fig:01} }
   \end{subfigure}  
    \begin{subfigure}[b]{0.49\linewidth}
    \includegraphics[width=\linewidth]{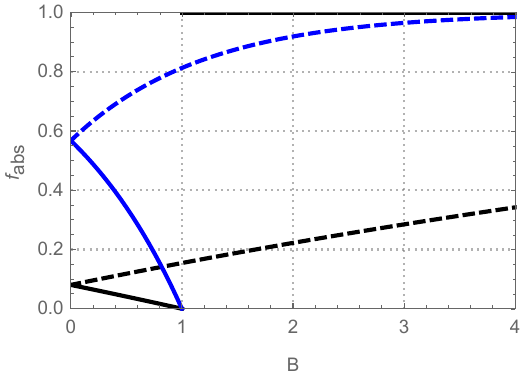}
    \caption{\label{fig:02} }
   \end{subfigure}  
    \begin{subfigure}[b]{0.48\linewidth}
    \includegraphics[width=\linewidth]{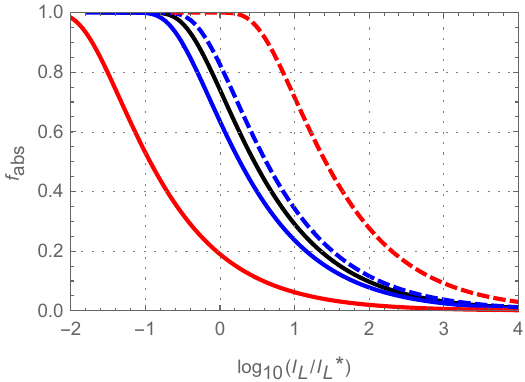}
    \caption{\label{fig:03} }
   \end{subfigure}  
    \begin{subfigure}[b]{0.50\linewidth}
    \includegraphics[width=\linewidth]{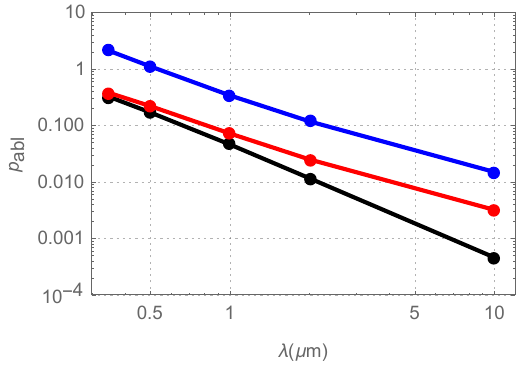}
    \caption{\label{fig:04} }
   \end{subfigure}  
  \captionsetup{justification=raggedright,singlelinecheck=false}
  \caption{\label{fig:absorption1} Absorption coefficient $f_{abs}$ of R wave (solid curve) and L wave (dashed curve), obtained from standing wave model (a,b) and travelling wave model  (c,d).  (a) $f_{abs}$ v.s $L$: $B=0$ (black), $B=0.2$ (blue), $B=0.8$ (red)  and $B=2$ (red), with $\nu_c=0.001$, $L_0=5\mu{m}$; (b) $f_{abs}$ v.s $B$: $L_0=5\mu{m}$ (black) and $L_0=50\mu{m}$ (blue), with $\nu_c=0.001$; (c)  $f_{abs}$ v.s $I_L/I_L^*$:  $B=0$ (black), $B=0.2$ (blue), $B=0.8$ (red) and $B=2$ (red); (d)  $p_{abl}(a.u.)$  v.s $\lambda(\mu{m})$ for L wave: $B_0=0$, $I_L= 10^2 *I_{L0}^* $ (black); $B_0=6*10^3$ T, $I_L= 10^2 *I_{L0}^* $ (red); $B_0=6*10^3$ T, $I_L= 10^4 *I_{L0}^* $ (blue); $I_{L0}^*= 2.6\times 10^{9} Z ln \Lambda\cdot L_0$. }
\end{figure*}

Travelling wave model is employed to obtain the absorption properties accounting for the plasma heating effect, which takes place mainly in region close to R- and L- cutoff, with whistler wave being excluded. The coefficient of collisional absorption is written as
\begin{align}
&f_{abs} = 1-exp\left[-2\int_{z_{min}}^{z_{max}} K(z)dz \right]
\end{align}
Based on Eq .\eqref{eq:K} and for plasma with constant density gradient, it becomes
\begin{align}
&f_{abs} \approx 1-exp\left[-\frac{32}{15}L(1\pm B)\nu_c \right]
\end{align}
where $\nu_c$ is also dependent on $f_{abs}$ due to the increased $T_{eV}$ in the absorption process, which is deduced here. As laser deposits energy in region close to cutoff with mass density of $\rho_{cut}=(1\pm B)m_i n_c$, the heated plasma will flow with velocity equal to local sound velocity $c_s=(T_{e}/m_i)^{1/2}$ \cite{batani2014physics}, where the energy conservation between absorbed laser energy flux and plasma flow is written as $f_{abs} I \approx \rho_{cut} c_s^3$. In this way, the plasma temperature scales as $T_{e}\approx m_i (f_{abs} I_L/4\rho_{cut})^{2/3}$ and thus the dimensionless electron-ion collisional frequency becomes (in practical units)
\begin{align}
&  \nu_c \approx 1.72 ln \Lambda Z \lambda_0^{-1}\cdot \frac{4\rho_{cut}}{f_{abs} I_L} \cdot \left( \frac{m_0}{e}\right) ^{-3/2}
\end{align}
In this way, the collisional absorption coefficient, without hidden variables,  is
\begin{align}
& f_{abs}=1-exp\left(-(1\pm B)^2 \frac{I_{L}^*}{f_{abs} I_L} \right)
\label{eq:pressure}
\end{align}
where
\begin{align}
&I_L^* \approx 14.7  \left( \frac{e^3}{m_i} \right)^{1/2}\frac{n_{c0}L_0}{\lambda_L^4} \approx 2.6\times 10^{9} Z ln \Lambda \frac{L_0}{\lambda^{4}}  \;  W/cm^2
\end{align}
is a critical intensity. Here, both $L_0$ and $\lambda$ are in units of micrometers,  $ I_L^*\approx 1.3 \times 10^{10}~ W/cm^2$ for $L_0=5~\mu\text{m}$ and $\lambda=1~\mu\text{m}$, assuming $Zln \Lambda =1$. Similar results for un-magnetized plasma in Ref. \cite{atzeni2004physics} is obtained with $B=0$ in Eq. \eqref{eq:pressure}. The dependences of $f_{abs}$ on laser intensity  for different magnetic field strength are shown in Fig. \ref{fig:03}, showing decreased absorption at higher intensity and relative higher absorption with L waves.  The shock pressure from laser ablation in non-magnetized plasma \cite{batani2014physics}  is  $ p_{abl} \approx 0.4 \rho_c^{1/3} (f_{abs}I_L)^{2/3}$ \cite{batani2014physics}, which can be easily converted to pressure in magnetized plasma in our case. 
\begin{align}
& p_{abl} \approx 0.4 \rho_c^{1/3} (1 \pm B)^{1/3} (f_{abs}I_L)^{2/3}
\end{align}
It is found L wave can generate stronger ablation pressure.  The dependence of L-wave ablation pressure on wavelength with different magnetic field strength and laser intensities are shown in Fig. \ref{fig:04}.  It is found that $p_{abl}$ increases with laser intensity and the dimensionless strength of magnetic field, however decreases with laser wavelength. The shock pressure from $CO_2$ laser with magnetic field strength of 6000 tesla ($B=6$) is far less being comparable with that from Nd:YAG laser in non-magnetized plasma, showing that shock pressure does not favours laser with large wavelength even with the help of strong magnetic field.  

\section{Discussion and conclusion} 
\label{sec:discussion}
We have systematically investigated the normal-incidence field-aligned propagation and collisional absorption of laser light in strongly magnetized, inhomogeneous plasmas with the magnetic fields parallel to the propagation direction, and have obtained the complete equations of electric fields in both vacuum and plasma for the first time.  Analytical models and PIC simulations consistently show that magnetic fields significantly alter laser propagation and collisional absorption compared to unmagnetized plasma.  Key findings include:

\begin{enumerate}
\item \textbf{Mode-dependent propagation:} 
\begin{itemize}
\item LCP laser is red-shifted and reflects at $n = 1 + B$ with enhanced amplitude;
\item RCP laser ($B<1$) is red-shifted and reflects at $n = 1 - B$ with enhanced amplitude;
\item RCP laser ($B>1$) propagates into overdense plasmas without cutoff, amplitude decreased, blue-shifted, pulse compressed and total absorbed;
\item LP laser ($B>1$) is split into forward RCP and backward LCP laser at L-cutoff,  with splitting ratio and wave profile tuned by the polarization of incident laser.
\end{itemize}
\item \textbf{Magnetic field strength:} For L wave, absorption increases with magnetic field; for R wave ($B<1$), absorption decreases to zero with magnetic field; for R wave ($B>1$), totally absorbed;  
\item \textbf{Plasma scale length:} Absorption increases with larger scale length, modulated by magnetic field.
\end{enumerate}

In both astrophysical and laboratory plasma, the propagation of electromagnetic waves in inhomogeneous plasma along the magnetic field is a common phenomenon, i.e. the radio and x-ray waves in the magnetosphere of neutron star and magnetized cosmic plasma, the laser and x-ray waves in the plasma of magnetized fusion, and et al., which could be, at least qualitatively, understood with the above analytical equations \eqref{eq:plasma}.  One example is the  hypersonic vehicle telemetry blackout \cite{starkey2015hypersonic}, where the communication of radio wave becomes blackout as hypersonic vehicle re-entries the atmosphere and generates over-dense plasma sheath on the vehicles surface. One solution is the application of a strong magnetic field, thus the whistler wave can propagate through the over-dense plasma and continues the communication, as shown in Fig. \ref{fig:RCP2}. 

Another potential application is in fast ignition, where the fast electron beams are generated by the relativistic laser at around critical density of plasma and then propagate a long distance to deposit energy to the compressed fuel. One of the main bottlenecks of fast ignition is the low energy deposition efficiency due to the long propagation distance and large divergence angle of electron beam. People have worked on the application of whistler wave to overcome this problem and found that the relativistic laser can propagate to high density plasma and generate fast electrons with larger energy and conversion efficiency compared to non-magnetized plasma  \cite{yang2015propagation,gong2017enhancing}. In these paper, however, the strength of magnetic field is as high as $2\sim3 \times 10^4$ tesla due to the application of Nd:YAG laser. Based on Fig. \ref{fig:04}, it seems that magnetic field as "low" as 6000 tesla with relativistic $CO_2$ laser could realize same results, making these research much more realistic. 

In conclusion, we have bridged the classical unmagnetized plasma theory with magnetized scenarios, shown comprehensive understanding of laser propagation and collisional absorption in highly-magnetized inhomogeneous plasma. Based on these results, further fundamental researches such as resonant absorption and parametric instabilities in the interactions of laser with highly-magnetized plasma will be performed. Hopefully, with the advances in strong magnetic field and powerful far infra-red laser, these findings might have realistic impact on research of laboratory plasma, including applications in high energy density plasma and inertial confinement fusion.  

\section*{Acknowledgments}
The authors gratefully acknowledge the discussion with Professor Vladimir Tikhonchuk of Centre Lasers Intenses et Applications and University of Bordeaux. This work is supported by the National Natural Science Foundation of China Project (U2330123). 

\section*{Data availability}
The data that support the findings of this study are available from the corresponding author upon reasonable request.

\section*{Disclosures.}
The authors declare no conflicts of interest.

\small
\bibliographystyle{apsrev4-1}
\bibliography{reference}

\end{document}